\documentclass[twocolumn]{article}
\newcommand{\nc}{\newcommand}

\newcommand{\ra} {R_{A}}
\newcommand{\rb} {R_{B}}

\nc{\fau} {F(x;\mu)}
\nc{\expand}[3]{#1 & = & \lim_{ x \rightarrow 0^{#2}} 
\frac {\partial} {\partial #3} F(x;0)}
\nc{\expandd}[4] {#1 & = & \lim_{ \bar{x} \rightarrow 0^{#2}, \, \bar{y}
\rightarrow 0}  
\frac {\partial} {\partial #3} #4}

\newcommand{\fbarxy}{F(\bar{x},\bar{y};\bar{\mu})}

\newcommand{\eo} {\mathrm{P}_{0}}

\newcommand{\ori} {\left( \begin{array}{c} 0\\0 \end{array} \right)}
\newcommand{\va} {\left( \begin{array}{c} v_{Lx}\\v_{Ly} \end{array} \right)}
\newcommand{\vb} {\left( \begin{array}{c} v_{Rx}\\v_{Ry} \end{array} \right)}
\newcommand{\vab} {\left( \begin{array}{c} v_{x}\\v_{y} \end{array} \right)}
\newcommand{\ox} {\overline{x}}
\newcommand{\oy} {\overline{y}}
\newcommand{\ou} {\overline{\mu}}
\newcommand{\xybar} {\left( \begin{array}{c}
\ox \\ \oy \end{array} \right)}
\newcommand{\la} {\lambda _{1}}
\newcommand{\lb} {\lambda _{2}}

\newcommand{\tl} {\tau_{L}}
\newcommand{\tr} {\tau_{R}}
\newcommand{\dl} {\delta_{L}}
\newcommand{\dr} {\delta_{R}}
\newcommand{\ma} {\left(\! \begin{array}{cc}
\tl & 1 \\
-\dl & 0
\end{array}\! \right)}
\newcommand{\mb} {\left( \!\begin{array}{cc}
\tr & 1 \\
-\dr & 0
\end{array}\! \right)}

\nc {\dleft}{-(1+\delta)}
\nc {\dright} {(1+\delta)}
\nc {\fixax} {\frac {\mu }{ 1- \tl + \delta}}
\nc {\fixay} {\frac {-\delta \, \mu }{ 1- \tl + \delta}}
\nc {\fixbx} {\frac {\mu }{ 1- \tr + \delta}}
\nc {\fixby} {\frac {-\delta \, \mu }{ 1- \tr + \delta}}
\nc {\labl}{ \frac {1}{2} ( \tl \pm  \sqrt{\tl^{2} - 4 \delta} \, )}
\nc {\lal}{ \frac {1}{2} ( \tl +  \sqrt{\tl^{2} - 4 \delta} \, )}
\nc {\xbcoor} {x_{0} + \frac{\la}{\delta} \, y_{0}}
\nc {\fina} {\frac {\mu} {1 - a}}
\nc {\finb} {\frac {\mu} { 1- b}}

\nc {\ber} {\begin{eqnarray}}
\nc {\eer} {\end{eqnarray}}
\nc {\beq} {\begin{equation}}
\nc {\eeq} {\end{equation}}

\newcommand{\ux} {\left(\! \begin{array}{c} 1\\0 \end{array}\! \right)}

\newcommand{\xy} {\left(\! \begin{array}{c} x\\y \end{array}\! \right)}

\nc {\uml} {{\bf U}_L}
\nc {\umr} {{\bf U}_R}
\nc {\sml} {{\bf S}_L}
\nc {\smr} {{\bf S}_R}

\nc {\bl} {\bar{L^*}}
\nc {\br} {\bar{R^*}}

\newcommand{\td}[1]{\tilde{#1}}
\newcommand{\ghat}{g(\hat{x},\hat{y};\rho)}
\newcommand{\ftilde}{f(\td{x},\td{y};\rho)}

\newcommand{\tyo}{\td{y}_0}

\usepackage{rotating}
\usepackage{epsfig}
\usepackage{example}
\begin{document}

\title{ \Large {\bf Border Collision Bifurcations in\\ Two Dimensional
Piecewise Smooth
Maps}}
\author{Soumitro Banerjee $^{1}$ and Celso
Grebogi $^{2} $\\ $^1$ {\em \normalsize Department of Electrical
Engineering, Indian Institute of Technology, Kharagpur--721302, India} \\
$^2$
{\em \normalsize Institute for Plasma Research, Department of
Mathematics and 
Institute for Physical Science \&
Technology,}\\ {\em \normalsize University of Maryland, College Park, MD
20742, USA}}
\date{\normalsize \begin{quote} 
Recent investigations on the bifurcations in switching circuits have shown
that many atypical bifurcations can occur in piecewise smooth maps which
can not be classified among the generic cases like saddle-node, pitchfork
or Hopf bifurcations occurring in smooth maps. In this paper we first
present experimental results to establish the need for the development of
a theoretical framework and classification of the bifurcations resulting
from border collision. We then present a systematic analysis of such
bifurcations by deriving a normal form --- the piecewise linear
approximation in the neighborhood of the border. We show that there can be
eleven qualitatively different types of border collision bifurcations
depending on the parameters of the normal form, and these are classified
under six cases. We present a partitioning of the parameter space of the
normal form showing the regions where different types of bifurcations
occur. This theoretical framework will help in explaining bifurcations in
all systems which can be represented by two dimensional piecewise smooth  
maps.
 \end{quote}}
\maketitle

\section{Introduction}

Most studies in bifurcation theory have been done using smooth dynamical
systems like the H\'{e}non map, the Ikeda map and the pendulum equation.
In the class of non-smooth systems, maps with square root singularity have
been studied extensively \cite{Chin94,Chin95,Nusse94,Budd94} because of
their application in impact oscillators and other impacting mechanical
systems. On the other hand, piecewise smooth maps with finite one-sided
partial derivatives at the discontinuity have attracted relatively little
attention. Though the possibility of strange bifurcations like period-2 to
period-3 or period-2 to 18-piece chaotic attractor have been reported
\cite{Nusse92}, no systematic study has been made to categorize the
possible bifurcations in piecewise smooth maps. Such maps were considered
to be just a mathematical possibility as no physical system with these
characteristics was known. 

However in recent years there has been a discovery that a large class
of engineering systems, particularly the switching circuits used in power
electronics, yield piecewise smooth maps under discrete modeling, and
border collision bifurcations are quite common in such systems
\cite{pesc97,Guohui97}. This has provided motivation for the present study
whose objective is to systematically analyze all different kinds of
bifurcations that can occur in two dimensional piecewise smooth maps. 

We consider a general two-dimensional piecewise smooth map
$g(\hat{x},\hat{y};\rho)$ which depends on a single parameter $\rho$. Let
$\Gamma_\rho$, given by $\hat{x}=h(\hat{y},\rho)$ denote a smooth curve that
divides the phase plane into two regions $R_A$ and $R_B$.  The map is given by

\begin{equation} g(\hat{x},\hat{y};\rho) = \left\{ \begin{array}{ll}
g_1(\hat{x},\hat{y};\rho) \;\; \mbox{for}\;\; \hat{x},\hat{y} \in R_A ,\\
g_2(\hat{x},\hat{y};\rho) \;\; \mbox{for} \;\; \hat{x},\hat{y} \in R_B
\end{array} \right. \label{psm}
\end{equation}
It is assumed that the functions $g_1$ and $g_2$ are both continuous and have
continuous derivatives. The map $g$ is continuous but its derivative is
discontinuous at the line $\Gamma_\rho$, called the ``border''. It is further
assumed that the one-sided partial derivatives at the border are finite. We
study the bifurcations of this system as the parameter $\rho$ is varied.

If a bifurcation occurs when the fixed point of the map is in one of the
smooth regions $R_A$ or $R_B$, it is one of the generic types, namely, period
doubling, saddle-node or Hopf bifurcation.  But if a fixed point collides with
the borderline, there is a discontinuous jump in the eigenvalue of the
Jacobian matrix. In such a case an eigenvalue may not ``cross'' the unit
circle in a smooth way, but rather ``jumps'' over it as a parameter is varied
continuously. One therefore cannot classify the bifurcations arising from such
border collisions as those occurring for smooth systems where the eigenvalues
cross the unit circle smoothly. In this paper we develop a new classification
for border collision bifurcations.

The paper is organized as follows. In Sec.~2, we illustrate the problem with
the help of an example of switching circuit. In Sec.~3, the normal form is
derived. In Sec.~4, we analyse the border collision bifurcations occurring in
piecewise smooth maps. We present a partitioning of the parameter space of the
normal form exhibiting various kinds of border collision bifurcations.  We
conclude in Sec.~5.

\section{Examples of border collision bifurcations in a power electronic
circuit}

The subject of power electronics is concerned with high efficiency conversion
of electric power, from the form available at the power source, to the form
required by the specific appliance or load.  Power electronic technology is
increasingly finding application in the home and workplace: familiar examples
are domestic light dimmers, fluorescent lamp ballasts, battery chargers, and
switch-mode power supplies of all electronic appliances including the personal
computer.

In contrast with mainstream electronics, power electronics is characterized by
the use of electronic {\em switches} which operate in ``on'' or ``off'' state.
Since electrical power supplies can be either dc or ac, there are four basic
types of power converters: ac-dc, dc-ac, dc-dc and ac-ac. Here we will
consider one of the simplest but most useful of power converters --- the dc-dc
buck converter --- which is used to convert a dc input to a dc output at a
lower voltage.

\begin{figure}[tbh]
\begin{center}
\epsfig{figure=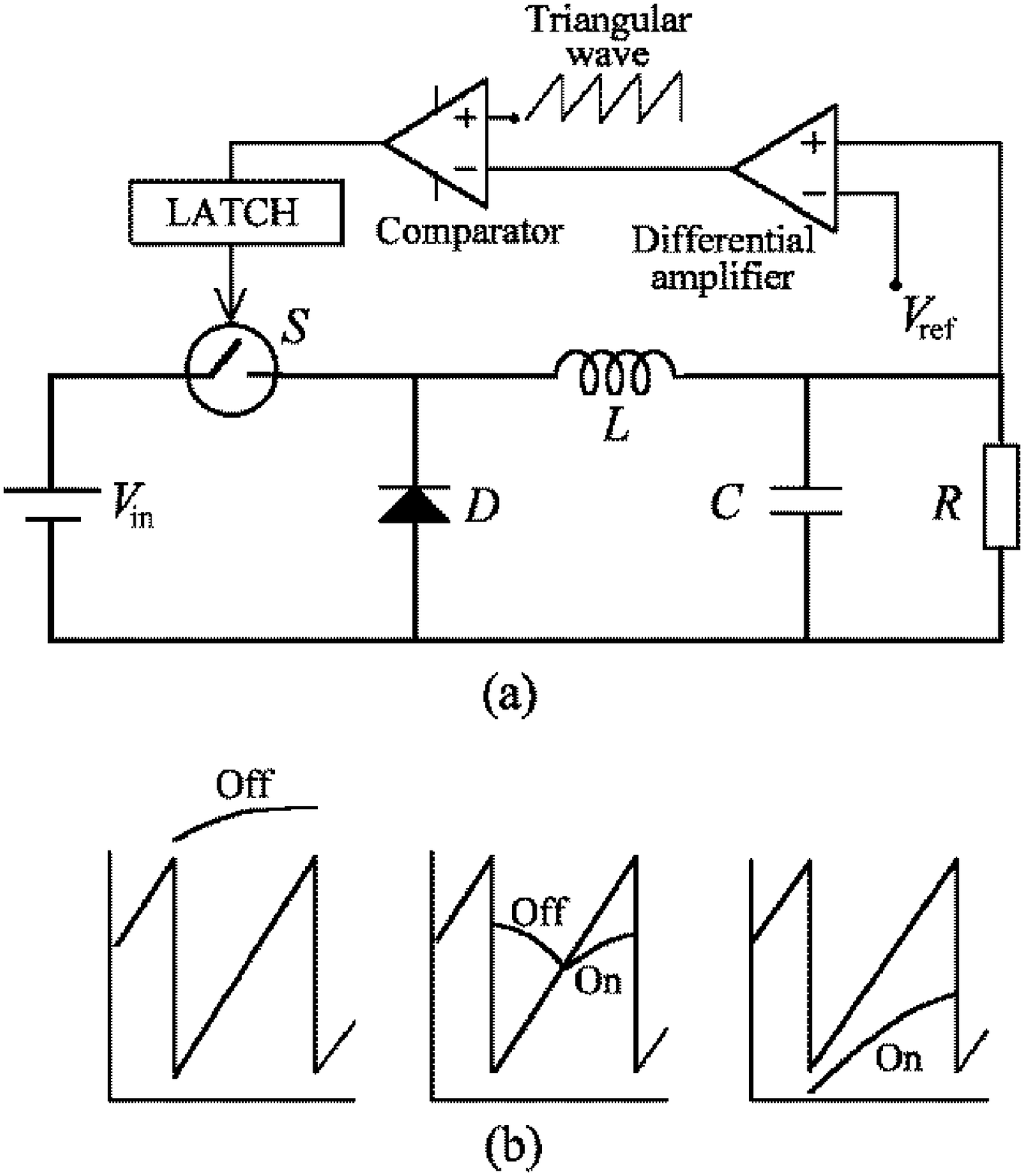,width=2.9in}\end{center}
 \caption{(a) The buck converter with duty cycle
controlled  
by voltage feedback, (b) The three ways the state can move from one
sampling instant to the next.} \label{vbuckfig} \end{figure}

The circuit diagram of the buck converter is shown in Fig.\ref{vbuckfig}(a).
The controlled switch $S$ (generally realized by a MOSFET) opens and closes in
succession, thus ``chopping'' the dc input into a square wave that alternates
between the input voltage $V_{in}$ and zero. The pulsed waveform is then
low-pass filtered by a simple $LC$ network, removing most of the switching
ripple and delivering a relatively smooth dc output voltage $v$ to the load
resistance $R$. The diode $D$ provides a path for the continuation of the
inductor current during the {\em off} period.  The dc output voltage can
easily be varied by changing the duty ratio, i.e., the fraction of time that
the switch is closed in each cycle.

In practice it is necessary to regulate $v$ against changes in the input
voltage and the load current. For example, if a buck converter is used to
convert the standard 5~V dc supply used in computers to the 3.3~V needed for
the Pentium CPU chip, it would be necessary to regulate the average output
voltage at 3.3~V in spite of the varying power demand of the chip. This can be
achieved by controlling the switch $S$ by voltage feedback as shown in
Fig.\ref{vbuckfig}.  In this simple proportional controller, a constant
reference voltage $V_{ref}$ is subtracted from the output voltage and the
error is amplified with gain $A$ to form a control signal
$v_{con}\!=\!A(v-V_{ref})$. The switching signal is generated by comparing the
control signal with a periodic sawtooth (ramp) waveform.  $S$ turns on
whenever $v_{con}$ goes below $v_{ramp}$ and a latch allows it to switch off
only at the end of the ramp cycle.

\begin{figure}[t]
\begin{center}
\epsfig{figure=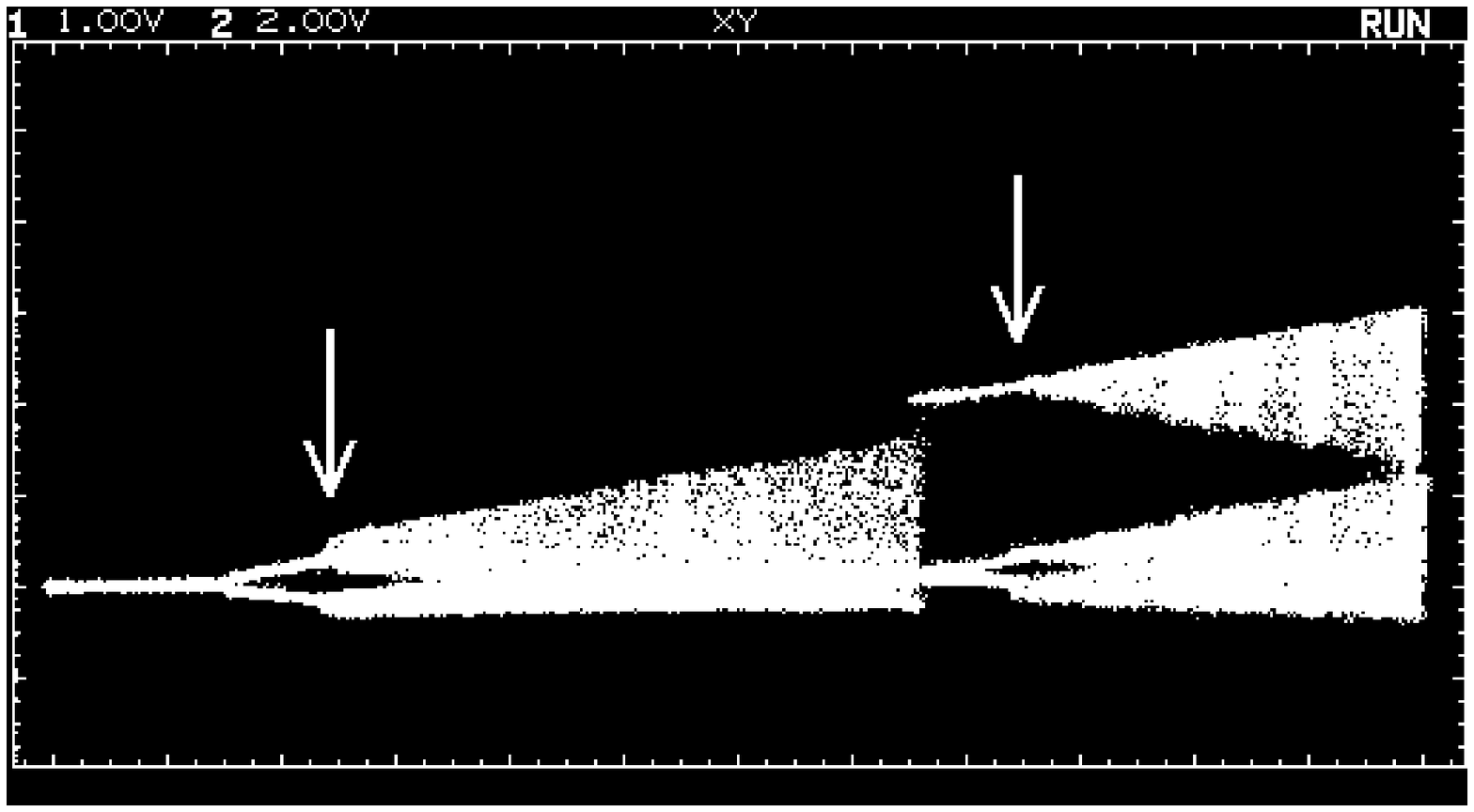,width=3in}(a)
\epsfig{figure=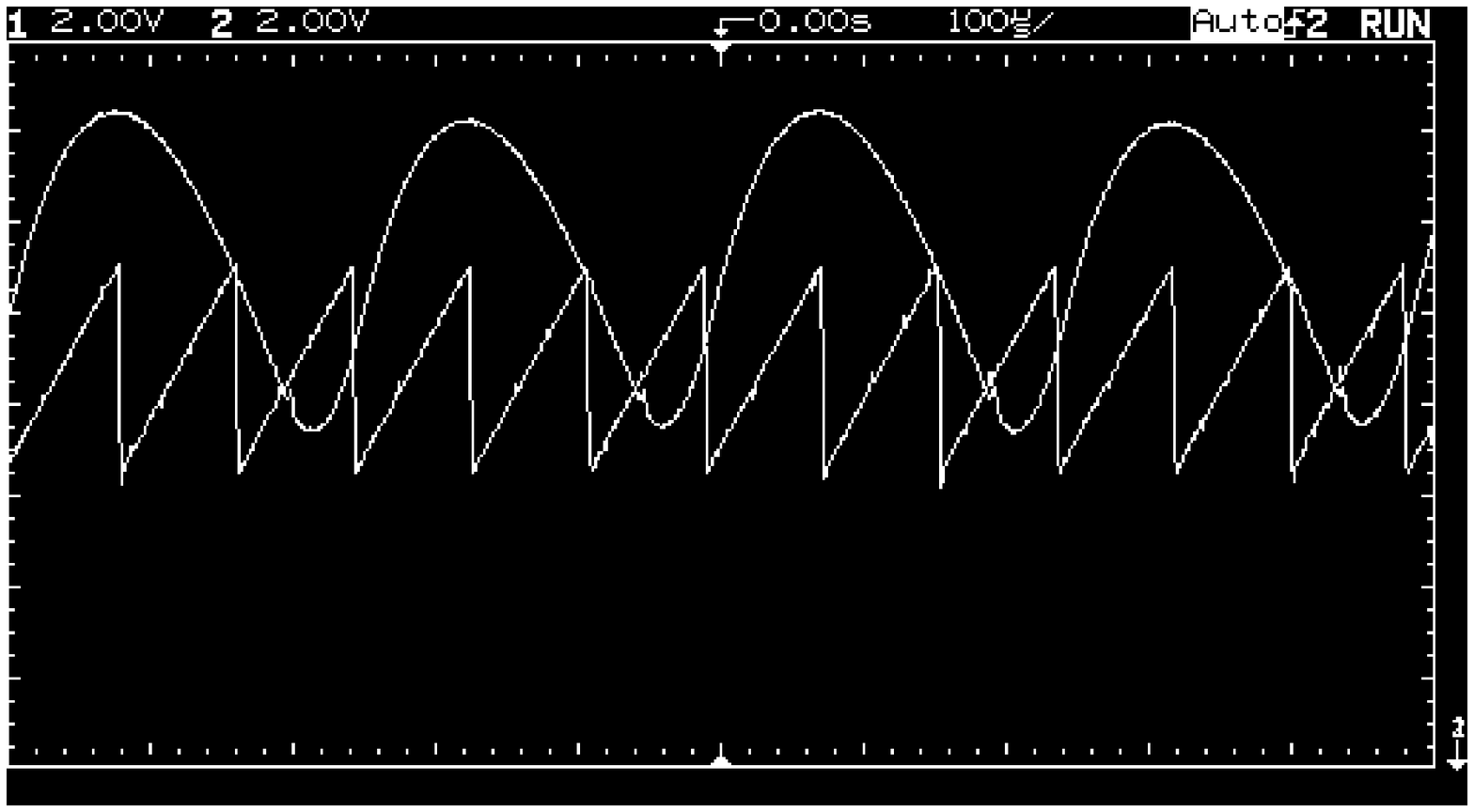,width=3in}(b)
\end{center}
\caption{Experimental bifurcation diagram of the buck converter. The   
parameter values are: $R=23.5\Omega$, $C=5\mu$F, $L=2.96$mH, triangular  
wave: $V_U=8.43$V, $V_L=3.62$V, frequency 12 kHz. Bifurcation parameter
$V_{in}$ varied from 35 to 75 V.}
\label{bif2}
\end{figure}

Though this circuit or its variants are used in a large number of practical
applications requiring regulated dc power supply, it has been demonstrated
\cite{DH90,Hamill95,HD92} that the system can exhibit
bifurcations and chaos for a large portion of the parameter space. To
investigate the dynamics analytically, we obtain a two dimensional
Poincar\'{e} map by sampling the inductor current and capacitor voltage at the
end of each ramp cycle. 

Because of the transcendental form of the equations, the map cannot be
determined in closed form. In simulation, the map has to be obtained
numerically. It is however possible to infer the form of the map.  There are
three ways in which the system can move from one observation point to the
next: (a) the control voltage is throughout above the ramp waveform and the
switch remains off, (b) the cycle involves an {\em off} period and an {\em on}
period, (c) the control voltage is throughout below the ramp waveform and the
switch remains on. The three cases are shown in Fig.\ref{vbuckfig}(b). These
are represented by three different expressions of the map. The borderlines
are given by the condition where the control voltage grazes the top and bottom
of the ramp waveform. Therefore there are three compartments in the phase
space, separated by two borderlines, and we have a piecewise smooth map.

\begin{figure}[t]
\begin{center}
\epsfig{figure=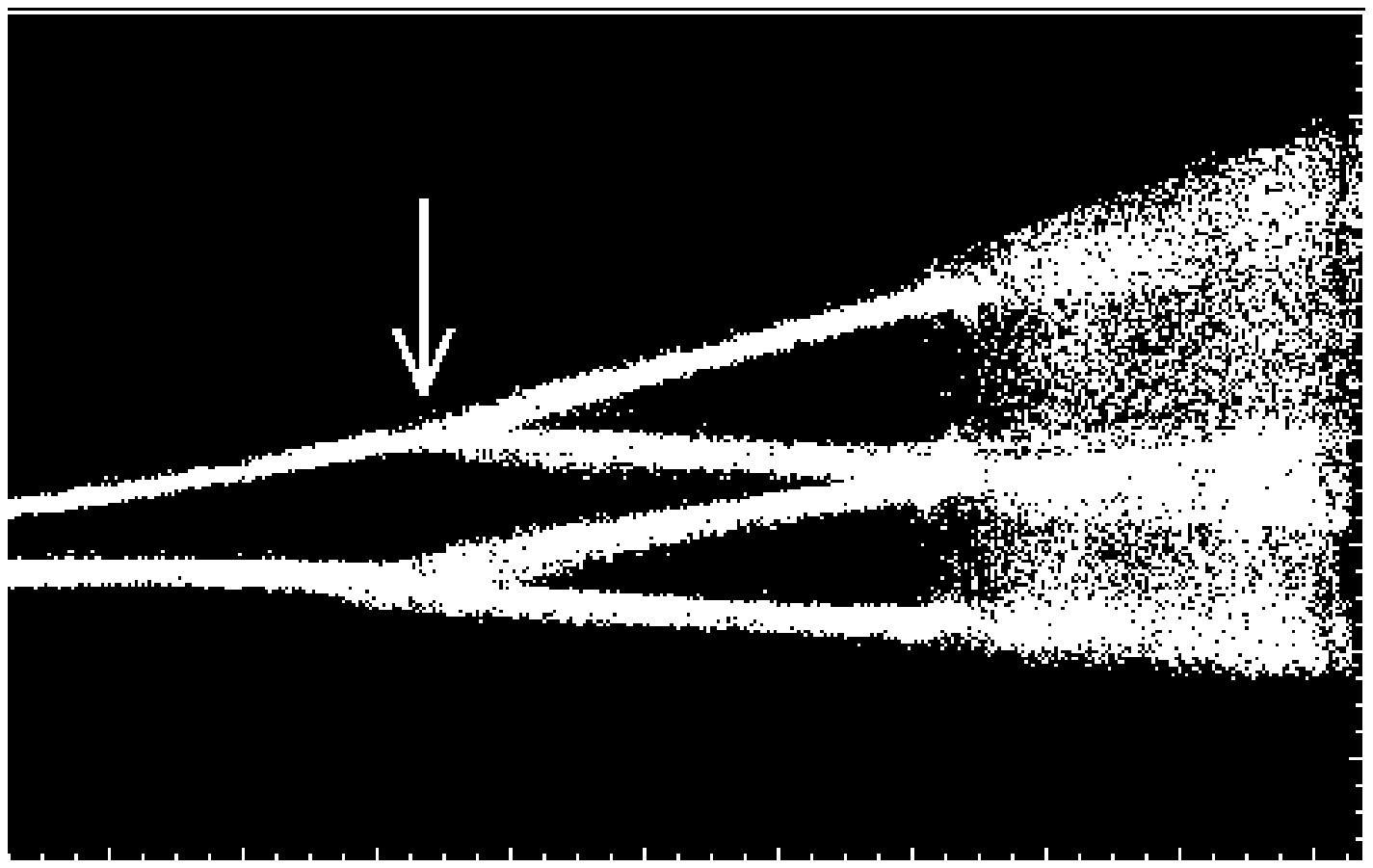,width=3in}(a)
\epsfig{figure=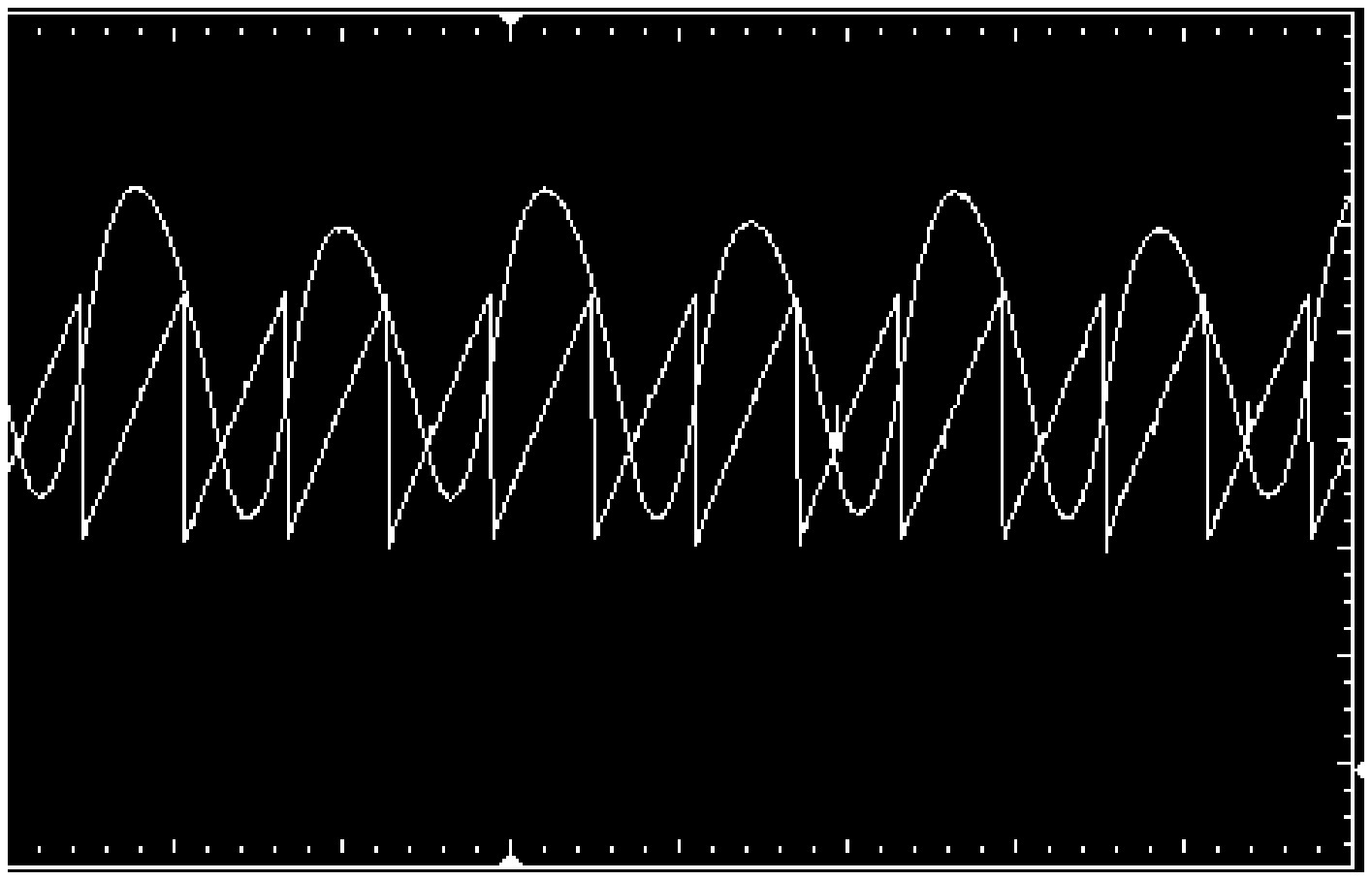,width=3in}(b)
\end{center}
\caption{Experimental
bifurcation diagram of the buck converter. The parameter values are:
$R=28.9\Omega$, $C=5\mu$F, $L=2.96$mH, triangular wave: $V_U=8.43$V,
$V_L=3.62$V, frequency 8 kHz. Bifurcation parameter $V_{in}$ varied from
50 to 70 V.}
\label{bbif1}
\end{figure}

We present the experimentally obtained bifurcation diagrams for this system
for different sets of parameter values.

An experimental bifurcation diagram is shown in Fig.\ref{bif2}(a). Here we
find two parameter values (shown with arrows) for which a periodic orbit
directly bifurcates into a chaotic orbit. Such bifurcations have been
reported earlier in \cite{DH90,TseChan95,OI94,ZY95}.  The slight
expansion
of the attractor at the bifurcation point is due to system noise and can
be ignored in theoretical studies. In Fig.\ref{bif2}(b) we present the
continuous time plots of $v_{con}$ and the triangular wave voltage at the
bifurcation point shown by the second arrow, where a period-3 orbit
bifurcates into a 3-piece chaotic orbit.  It is seen that the $v_{con}$
waveform grazes the top of the triangular wave, which means that a border
collision bifurcation has occurred. 

The distinguishing feature of this chaotic attractor is that there is no
periodic window over a large range of the parameter value. We find from
simulation that there are no coexisting attractors in this range. We say a
chaotic attractor is {\em robust} if, for its parameter values there
exists a neighborhood in the parameter space with no periodic attractor
and the chaotic attractor is unique in that neighborhood \cite{Robust}.
The chaotic attractor resulting from this border collision is therefore
robust. The question is, under what condition does robust chaos occur?

\begin{figure}[tbh]
\begin{center}
\epsfig{figure=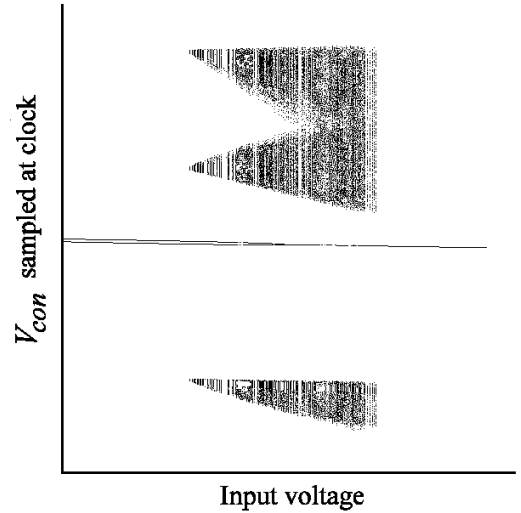,width=3.2in}   
\end{center}
\caption{Numerically obtained bifurcation diagram of the buck converter.
The parameter values are: $R=22\Omega$, $C=47\mu$F, $L=20$mH, triangular
wave: $V_U=8.2$V, $V_L=3.8$V, time period 400$\mu$s. }
\label{bbif3}
 \end{figure}

Another experimental bifurcation diagram for this system is shown in
Fig.\ref{bbif1}(a).  The arrow shows a period doubling bifurcation, but
the two bifurcated orbits do not diverge perpendicularly from the path of
the fixed point before the critical parameter value. This is therefore not
a standard pitchfork bifurcation. This kind of bifurcation has been
reported in \cite{Deane92,ChanTse96} also. Fig.\ref{bbif1}(b) gives the
continuous time plots of $v_{con}$ and the triangular wave voltage just
after the bifurcation and shows that the period doubling occurred at a
border collision. Again the question is, under what condition does this
special type of period doubling occur? 

It has been reported earlier \cite{SB96} that this system has coexisting
attractors for some ranges of parameter values. Since multiple attractors
cannot be seen in experimental bifurcation diagrams, we present a numerically
obtained bifurcation diagram in Fig.\ref{bbif3} showing the evolution of
the main attractor and a coexisting attractor. It is found that the chaotic
attractor comes into existence out of nothing at a particular parameter value.
Under what condition can such strange bifurcations occur?

In the following sections we develop a complete theory of bifurcations in
piecewise smooth maps, from which the answers to the above questions can be
derived.

\begin{figure*}[tbh]  
\begin{center}
\epsfig{figure=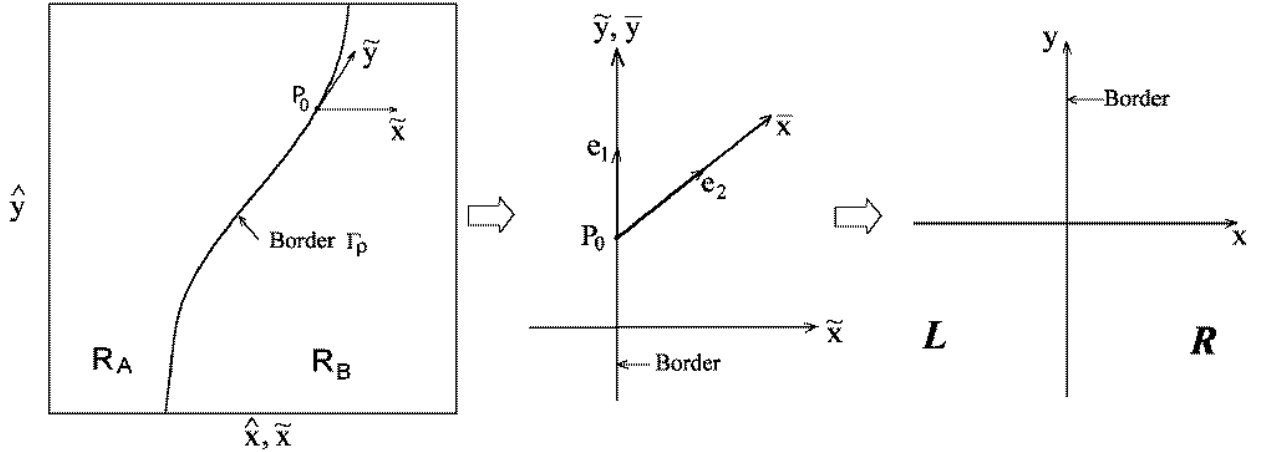,width=2.5in,angle=-90}
\end{center}
\caption{The transformation of coordinates from the two dimensional
piecewise smooth map to the normal form.}
\label{fig:frame}
\end{figure*}

\section{The normal form}

\label{norm-2d}

Since the local structure of border collision bifurcations depends only on the
local properties of the map in the neighborhood of the border, we study the
border collision bifurcations with the help of ``normal forms'' --- the
piecewise affine approximations of $g$ in the neighborhood of the border.

Define \[ \td{x}=\hat{x}-h(\hat{y};\rho) , \hspace{0.1in} \mbox{and}
\hspace{0.1in}  \td{y}=\hat{y}. \] This 
$\rho$-dependent change of variables moves the border to the $\td{y}$ axis.
Then the map $\ghat$ can be written
\[ g(\td{x}+h(\hat{y};\rho),\td{y};\rho)=\ftilde, \] 
and the border is $\td{x}=0$.  Suppose that when $\rho = \rho_0$ the map
$\ftilde$ has a fixed point $\eo$ on the border, that is, \[ \mbox{$\eo$} =
(0,\tyo(\rho_0)) = f(0,\tyo(\rho_0);\rho_0). \] Let $e_1$ be a tangent vector
in the $\td{y}$ direction. The vector $e_{1}$ maps to a vector $e_{2}$.  We
assume $e_2$ is not parallel to $e_1$. Define the local coordinates as the
following (C.f. Fig.~\ref{fig:frame}). Choose the point $\eo$ as the new
origin for $e_1$ in the $\bar{y}$ direction and $e_2$ in the $\bar{x}$
direction. In these $\bar{x}$-$\bar{y}$ coordinates, the fixed point $\eo$ is
given by $(0,0)$, and the border $\Gamma_\rho$ is given by $\bar{x}=0$. We
define the new parameter $\bar{\mu}=\rho-\rho_0$ so that $\bar{\mu}_0 = 0$.
Choose the scales such that at $\bar{\mu}\!=\!0$ an unit vector along
$\bar{y}$-axis maps to an unit vector along $\bar{x}$-axis. The phase space is
now divided into the two halves $L$ and $R$ and the map $\ftilde$ can be
written as $F(\bar{x},\bar{y};\bar{\mu})$.

We can write the map $\fbarxy$ in the side $L$ in the matrix form as
\[  \fbarxy = \left( \begin{array}{cc} 
f_{1}(\bar{x},\bar{y};\bar{\mu}) \\ f_{2}(\bar{x},\bar{y};\bar{\mu})
\end{array} \right), \;\;\;
\mbox{and} \;\;\;  F(0,0;0) = \ori. \]

Linearizing $\fbarxy$ in the neighborhood of (0,0;0), we have:

\begin{eqnarray} \fbarxy &=& {\left( \begin{array}{cc} J_{11} & J_{12} \\ 
J_{21} & J_{22} \end{array} \right)} \left( \begin{array}{c}
\bar{x}\\\bar{y} \end{array} \right) + \bar{\mu} \va +
\nonumber \\ && o(\bar{x},\bar{y};\bar{\mu}), \;\;\; \mbox{for $\bar{x}
\leq 0$ }
\label{matrix} \end{eqnarray} where \begin{eqnarray*}
\expandd{J_{11}}{-}{\bar{x}}{f_{1}(\bar{x},\bar{y};0)}, \\
\expandd{J_{12}}{-}{\bar{y}}{f_{1}(\bar{x},\bar{y};0)}, \\
\expandd{J_{21}}{-}{\bar{x}}{f_{2}(\bar{x},\bar{y};0)}, \\
\expandd{J_{22}}{-}{\bar{y}}{f_{2}(\bar{x},\bar{y};0)}, \\
\expandd{v_{Lx}}{-}{\bar{\mu}}{f_{1}(\bar{x},\bar{y};0)}, \\
\expandd{v_{Ly}}{-}{\bar{\mu}}{f_{2}(\bar{x},\bar{y};0)}.  \end{eqnarray*}

Since an unit vector along $\bar{y}$-axis
maps to an unit vector along $\bar{x}$-axis, by (\ref{matrix}) this
particular choice of coordinates makes
$J_{12}=1$ and $J_{22}=0$. Further,
we note that $J_{11}$ is the trace (denoted $\tl$) and $J_{21}$ is the
negative of the determinant (denoted $-\dl$) of the Jacobian matrix. Thus
(\ref{matrix}) becomes

\begin{eqnarray} \label{eq:expand2d}
F(\ox,\oy;\ou) &=& \ma \xybar + \ou \left( \begin{array}{c} v_{Lx}\\v_{Ly}
\end{array} \right)\nonumber \\ &&+ o(\ox,\oy;\ou), \;\;\; \mbox{if
$\;\;\;
\ox \leq
0$,} \end{eqnarray} Similarly, for side $R$ we obtain
\begin{eqnarray}
F(\ox,\oy;\ou) &=& \mb \xybar + \ou \left( \begin{array}{c} v_{Rx}\\v_{Ry}
\end{array} \right)\nonumber \\ && + o(\ox,\oy;\ou), \;\;\; \mbox{if
$\;\;\;
\ox > 0$.}
\end{eqnarray}
where the corresponding quantities in $R$ are defined in a similar way.

Continuity of the map implies
\[	\va = \vb = \vab . \] 

We now make another change of variables so that the choice of axes is
independent of the parameter. The coordinate transformation $x = \ox$, and $y
= \oy - \bar{\mu} \, v_{y}$, and $\mu = \ou \, (v_x + v_y)$ (assuming $(v_x +
v_y) \neq 0$) gives

\begin{eqnarray} \label{normal}
G_{2} = \left\{ \begin{array}{ll} \ma \xy + \mu \ux, & \mbox{for} \;
x \leq 0, \\ \mb \xy + \mu \ux, & \mbox{for} \; x > 0, \end{array} \right.
\end{eqnarray}
which is the desired 2-D normal form.

Note that if $(v_x + v_y) = 0$, then the fixed point moves along the border as
$\mu$ varies. Hence we assume the genericity condition $(v_x + v_y) \neq 0$ to
ensure that a border collision occurs at $\mu=0$.

It is interesting to note that $\tl$ and $\dl$ are simply the trace and the
determinant of the Jacobian matrix of the fixed point $\eo$ on $\ra$ side of
the border $\Gamma$.  Let $P_{\rho}$ denote a fixed point of
$g(\hat{x},\hat{y};\rho)$ defined on $\rho_0-\epsilon <
\rho < \rho_0+\epsilon$ for some small $\epsilon >0$, then $P_{\rho}$ depends 
continuously on $\rho$. Assume that $P_{\rho}$ is in region $\ra$ when
$\rho\!<\!\rho_0$ and in region $\rb$ when $\rho\!>\!\rho_0$, and that
$P_{\rho}$ is on $\Gamma$ when $\rho\!=\!\rho_0$. For $\rho\!<\!\rho_0$, the
eigenvalues of the Jacobian matrix of the fixed point $P_{\rho}$ are denoted
as $\la$ and $\lb$. Since the trace and the determinant of the Jacobian is
invariant under the transformation of coordinates, we can obtain the values of
$\tl$ and $\dl$ as
\begin{eqnarray}
\tl & = & \lim_{\rho \rightarrow \rho_0^{-}}( \la + \lb ) \nonumber \\ \dl & = &
\lim_{\rho \rightarrow \rho_0^{-}} \; (\la \, \lb ). 
\label{eq:eigen}
\end{eqnarray}
The values of $\tr$ and $\dr$ can be calculated in a similar way for
$\rho\!>\!\rho_0$.  This property is very important in numerical computations.
For a border-crossing periodic orbit with higher period, we examine the $p$th
(if the period is $p$) iterate of the map.  The matrices in (\ref{normal})
then correspond to the $p$th iterate rather than the first iterate of the map.

When $\delta_L$ and $\delta_R$ are zero, the system becomes one dimensional
and the normal form reduces to
\begin{equation} \label{eq:1dnorm}
G_{1}(x;\mu) = \left\{ \begin{array}{ll} a \; x + \mu & \mbox{for $x \leq 0,$}
\\
b \; x + \mu & \mbox{for $x > 0 ,$} \end{array} \right.
\end{equation}
where $a$ and $b$ are the slopes of the graph at the two sides of the border
$x\!=\!0$.

\section{Classification of border collision bifurcations}

Various combinations of the values of $\tl$, $\tr$, $\dl$ and $\dr$ exhibit
different kinds of bifurcation behaviors as $\mu$ is varied through zero. To
present a complete picture, we break up the four dimensional parameter space
into regions with the same qualitative bifurcation phenomena.  If the
parameter combination is inside a region, then $g$ and $G_2$ will have the
same types of bifurcations. If it is on a boundary, then higher order terms
are needed to determine the bifurcations of $g$.

The fixed points of the system in both sides of the boundary are given by
\begin{eqnarray*}
L^* &=& \left( \frac{\mu}{1-\tl+\dl},\frac{-\dl \mu}{1-\tl+\dl} \right) \\ R^*
&=& \left( \frac{\mu}{1-\tr+\dr},\frac{-\dr \mu}{1-\tr+\dr} \right)
\end{eqnarray*}
and the stability of each of them is determined by the eigenvalues
$\lambda_{1,2}\!=\!\frac{1}{2}\left( \tau \pm \sqrt{\tau^2-4\delta} \right) $.
If the eigenvalues are real, the slopes of the corresponding eigenvectors are
given by $-(\delta / \lambda_1)$ and $-(\delta / \lambda_2)$, respectively.
Since we consider only dissipative systems, we assume $|\dl|<1$ and $|\dr|<1$.
Under this condition there can be four types of fixed points.

\begin{enumerate}

\item When $\delta > \tau^{2}/4$, both eigenvalues of the Jacobian are
complex, indicating that the fixed point is spirally attracting. If
$\tau>0$, it is a clockwise spiral, and if $\tau < 0$ the spiralling
motion is counter-clockwise. 

\item When $\delta < \tau^{2}/4$, both eigenvalues are real. If $2
\sqrt{\delta} < \tau < (1+\delta)$ then the eigenvalues are positive and
the fixed point is a regular attractor. If $-2 \sqrt{\delta} > \tau >
-(1+\delta)$ then the eigenvalues are negative and  it is a flip attractor.

\item  If $ \tau > 1+ \delta$, then $0 < \lambda_2 <1$ and
$\lambda_1> 1$. The fixed point is a regular saddle. 

\item  If
$\tau<-(1+\delta)$,
then $\lambda_2 < -1$ and $ -1<\lambda_1<0$. The fixed point is a flip
saddle.

\end{enumerate}

If the determinant is negative,
there can be only two types of
fixed points:   
\begin{enumerate}

\item For $-(1+\delta )\!<\!\tau
\!<\!(1+\delta )$, one eigenvalue is positive and the other negative ---
which means that the fixed point is a flip attractor.
\item For
$\tau>(1+\delta)$, $\lambda_1>1$ and $-1<\lambda_2<0$, i.e., the fixed
point is a flip saddle. If $\tau\!<\!-(1+\delta )$, then $\lambda_2 < -1$
and $0 < \lambda_1 <1$. The fixed point is again a flip saddle.
\end{enumerate}

When referring to sides $L$ and $R$, these quantities have the appropriate
subscripts, i.e., $\lambda_{1L}$, $\lambda_{2L}$ are the eigenvalues in side
$L$ and $\lambda_{1R}$, $\lambda_{2R}$ are the eigenvalues in side $R$.  As a
fixed point collides with the border, its character can change from any one of
the above types to any other. This provides a way of classifying border
collision bifurcations.

It may be noted that in some portions of the parameter space there may be no
fixed point in half of the phase space. For example, the location of $L^*$
calculated by the above formula may turn out to be in side $R$. In such cases,
the dynamics in $L$ is determined by the character of the ``virtual'' fixed
point. We denote such virtual fixed points by the overbar sign, as $\bar{L^*}$
and $\bar{R^*}$. If the eigenvalues are real, invariant manifolds of these
virtual fixed points still exist and play an important role in deciding the
system dynamics.

It should also be noted that if a certain kind of bifurcation occurs when
$\mu$ is increased through zero, the same kind of bifurcation would also occur
when $\mu$ is decreased through zero if the parameters in $L$ and $R$ are
interchanged.  Therefore, there exists a symmetry in the parameter space and
in the following discussion it suffices to describe the bifurcations in half
the parameter space. Moreover, we first consider the case of positive
determinant, which constitute a large class of physical systems. We take up
the special features of systems with negative determinant at a later stage.

A special feature of the normal form (\ref{normal}) is that the unstable
manifolds fold at every intersection with the x-axis, and the images of every
fold point is a fold point. The stable manifolds fold at every intersection
with the y-axis and the pre-image of every fold point is a fold point. The
argument is as follows.  Forward iterate of points on the unstable manifold
remain on the same manifold. In the normal form, points on the y-axis map to
points on the x-axis. As an unstable manifold crosses the y-axis, one linear
map changes to another linear map. Therefore the slope of the unstable
manifold in the two sides of the x-axis cannot be the same unless the
parameters of the normal form in the two sides of the border are the same
(implying a smooth map).  In case of the stable manifold, the same argument
applies for the inverse map.  Under the action 
of $G_2$, the line $x=0$ maps to the line $y=0$. Therefore under the
action of $G_2^{-1}$, points on the x-axis map to points on the y-axis,
and hence the stable manifold must have different slopes in the two sides
of the y-axis.

\begin{figure*} [tbh]
\begin{center}
\epsfig{figure=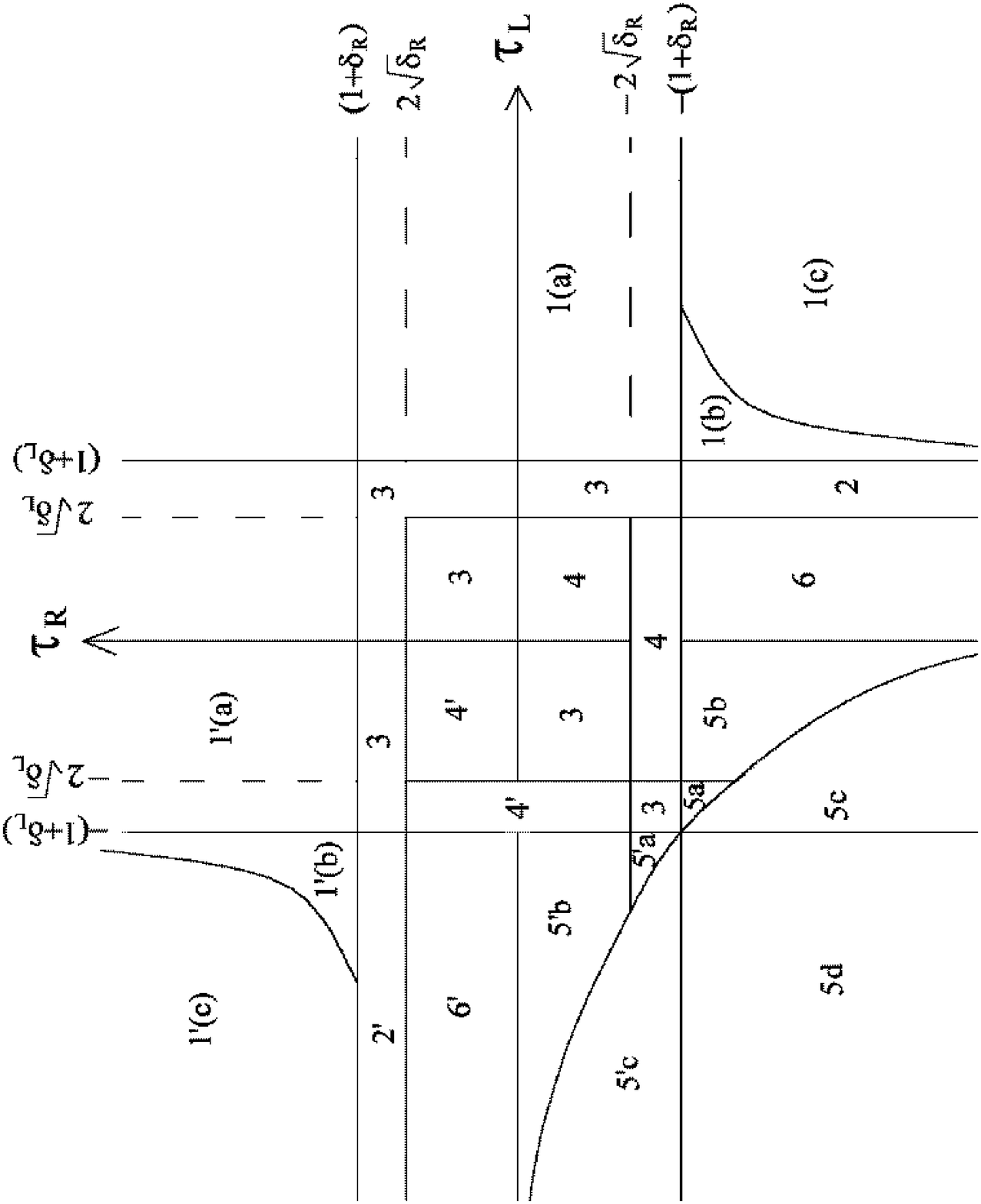,width=4.3in,angle=-90} 
\end{center}
\caption{The
partitioning of the parameter space into regions with the same qualitative
bifurcation phenomena. The numbering of the regions are the cases as
discussed in the text. The regions shown in primed numbers have the same
bifurcation behavior as the unprimed ones when $\mu$ is varied in the
opposite direction. }
\label{paramap}
\end{figure*}

We now present the partitioning of the parameter space as shown in
Fig.\ref{paramap}.  The system behavior in the various regions of the
parameter space are taken up in the following subsections.

\subsection{Border collision pair bifurcation}

\noindent {\bf Case 1:} 
If \beq \tl \!>\!(1+\dl) \;\;\; \mbox{and} \;\;\;\tr
\!<\!(1+\dr), \label{pair} \eeq then there is no fixed point for $\mu\!<\!0$
and there are two fixed points, one each in $L$ and $R$, for $\mu\!>\!0$. The
two fixed points are born on the border at $\mu\!=\!0$. We call this a {\em
border collision pair} bifurcation. An analogous situation occurs if $\tl
\!<\!(1+\dl)$ and $\tr \!>\!(1+\dr)$ as $\mu$ is reduced through zero. Due to
the symmetry of the two cases, we consider only the parameter region
(\ref{pair}).  There can be three types of border collision pair bifurcations
depending on the character of the orbits for $\mu>0$.

\noindent {\bf Case 1(a):}
If $(1+\dr)\!>\tr\!>-(1+\dr)$, then $R^*$ is stable. Therefore it is like a
saddle-node bifurcation, where a periodic attractor appears at $\mu\!=\!0$.
There are two special features of this saddle node bifurcation. First, the
fixed points are born on the border and move away from it as $\mu$ is
increased. Second, there is no intermittency associated with this bifurcation.

\noindent {\bf Case 1(b):} If

\beq \tl \!>\!(1+\dl) \;\;\; \mbox{and} \;\;\;\tr \!<\!-(1+\dr) \;\;\; 
\mbox{and} \label{pair2} \eeq \beq \dl \tr \lambda_{1L} - \dr \lambda_{1L}
\lambda_{2L} + \dr \lambda_{2L} - \dl \tr + \tl\dl - \dl^2 -
\lambda_{2L}\dl >0 \label{cond} \eeq there is a bifurcation from no
attractor to a chaotic attractor.  The chaotic attractor for $\mu>0$ is
robust \cite{Robust}. 

\noindent{\bf Case 1(c):} If $\tl \!>\!(1+\dl)$ and $\tr \!<\!-(1+\dr)$
and
\[\dl
\tr \lambda_{1L} - \dr \lambda_{1L} \lambda_{2L} + \dr \lambda_{2L} - \dl
\tr + \tl\dl - \dl^2 - \lambda_{2L}\dl \leq 0 \] then there is an unstable
chaotic orbit for $\mu\!>\!0$.  

For (\ref{pair2}), $L^*$ is a regular saddle and $R^*$ is a flip saddle. Let
$\uml$ and $\sml$ be the unstable and stable manifolds of $L^*$ and $\umr$ and
$\smr$ be the unstable and stable manifolds of $R^*$, respectively. As shown
earlier, $\uml$ and $\umr$ experience folds along the x-axis, and all images
of fold points are fold points. $\sml$ and $\smr$ fold along the y-axis, and
all pre-images of fold points are fold points.

For condition (\ref{pair2}), $\lambda_{1L}\!>\!\lambda_{2L}\!>\!0$ and
$0\!>\!\lambda_{1R}\!>\!\lambda_{2R}$.  The stable eigenvector at $R^*$ has a
slope $m_1\!=\!(-\dr/\lambda_{1R})$ and the unstable eigenvector has a slope
$m_2\!=\!(-\dr/\lambda_{2R})$. Since points on an eigenvector map to points on
the same eigenvector and since points on the y-axis map to the x-axis, we
conclude that points of $\umr$ to the left of y-axis map to points above
x-axis. From this we find that $\umr$ has an angle $m_3\!=\!(\dl
\lambda_{2R})/(\dr-\tl \lambda_{2R})$ after the first fold. Under condition
(\ref{pair2}) we have $m_1\!>\!m_2\!>\!0$ and $m_3\!<\!0$. Therefore there
must be a transverse homoclinic intersection in $R$. This implies an infinity
of homoclinic intersections and the existence of a chaotic orbit.

\begin{figure} [tbh]
\begin{center}
\epsfig{figure=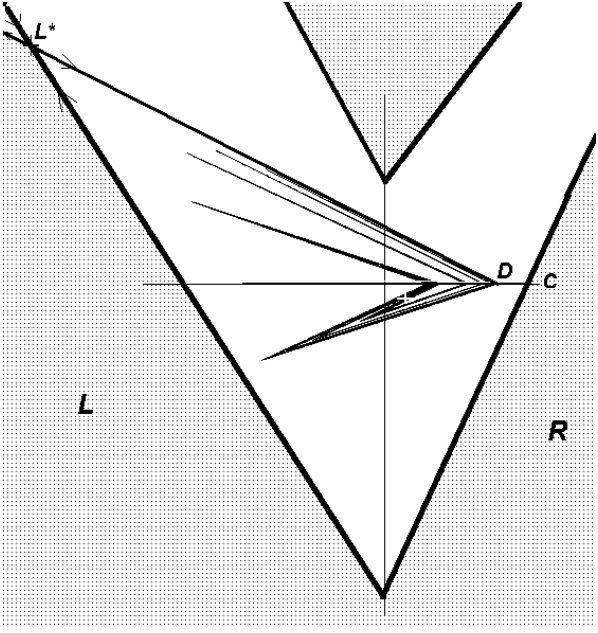,width=3.1in}\end{center}          
\caption{The
stable and unstable manifolds of $L^*$ for $\tl\!=\!1.7$, $\dl\!=\!0.5$,
$\tr\!=\!-1.7$, $\dr\!=\!0.5$. $R^*$ is marked by the small cross inside
the attractor.} \label{trap} \end{figure}

We now investigate the stability of this orbit. The basin boundary is formed
by $\sml$. $\sml$ folds at the y-axis and intersects the x-axis at point $C$.
The portion of $\uml$ to the left of $L^*$ goes to infinity and the portion to
the right of $L^*$ leads to the chaotic orbit.  $\uml$ meets the x-axis at
point $D$, and then undergoes repeated foldings leading to an intricately
folded compact structure as shown in Fig.\ref{trap}.

The unstable eigenvector at $L^*$ has a negative slope given by
$(-\dl/\lambda_{1L})$. Therefore it must have a heteroclinic intersection with
$\smr$. Since both $\uml$ and $\umr$ have transverse intersections with
$\smr$, by the Lambda Lemma \cite{yorkebook} we conclude that for each point
$q$ on $\umr$ and for each $\epsilon$-neighborhood $N_\epsilon (q)$, there
exist points of $\uml$ in $N_\epsilon (q)$. Since $\uml$ comes arbitrarily
close to $\umr$, the attractor must span $\uml$ in one side of the
heteroclinic point.

Since all initial conditions in $L$ converge on $\uml$ and all initial
conditions in $R$ converge on $\umr$, and since there are points of $\uml$ in
every neighborhood of $\umr$, we conclude that the attractor is unique. This
chaotic attractor cannot be destroyed by small changes in the parameters.
Since small changes in the parameters can only cause small changes in the
Lyapunov exponents, where the chaotic attractor is stable, it is also robust.

It is clear from this geometrical structure that no point of the attractor can
be to the right of point $D$.  If $D$ lies towards the left of $C$, the
chaotic orbit is stable. If $D$ falls outside the basin of attraction, it is
an unstable chaotic orbit or chaotic saddle. From this, the condition
(\ref{cond}) of stability of the chaotic attractor is obtained.  If
$\dl\!=\!\dr\!=\!\delta$ this condition reduces to
$\tr\lambda_{1L}-\lambda_{1L}\lambda_{2L}+\tl-\tr-\delta\!>\!0 $.

\subsection{Border crossing bifurcations}

In all regions of the parameter space except (\ref{pair}), a fixed point
crosses the border as $\mu$ is varied through zero. The resulting bifurcations
are called {\em border crossing} bifurcations. In the following discussions we
consider the bifurcations as $\mu$ varies from a negative value to a positive
value.

\noindent {Case 2:} {\em Linear attractor to flip saddle.} This occurs \[
\mbox{if} \hspace{0.1in} 2\sqrt{\dl} < \tl < (1+\dl)  \hspace{0.1in}
\mbox{and} \hspace{0.1in} \tr < -(1+\dr). \] There is a bifurcation from a
period-1 attractor to a chaotic attractor as $\mu$ is increased through
zero. This chaotic attractor is robust. 

For $\mu<0$, $L^*$ is a linear attractor while $\bar{R^*}$ is a flip saddle.
All initial conditions in $L$ converge on to $L^*$, while initial conditions
in $R$ converge on to $\umr$. Since $\umr$ must have a heteroclinic
intersection with one of the stable manifolds of $L$, all initial conditions
in $R$ also converge on to $L^*$.

For $\mu>0$, $R^*$ is a flip saddle. As shown in the discussion for Case~1(b),
there is a homoclinic intersection in $R$ implying the existence of a chaotic
orbit. As $\bl$ is in $R$, its stable manifolds point towards $R$. Since there
is an intersection of $\smr$ with the invariant manifold associated with
$\lambda_{1L}$, all initial conditions converge on $\umr$, making the chaotic
attractor unique.

\noindent {\bf Case 3:} There is a unique period-1 attractor for both
positive and
negative values of $\mu$ in the following cases. At border collision, only
the path of the fixed point changes. 

 {\em Regular attractor to spiral attractor:} This occurs 
\[\mbox{if} \;\;\;\; 2\sqrt{\dl} < \tl < (1+\dl) \hspace{0.1in} \mbox{and}
\hspace{0.1in} -2\sqrt{\dr} < \tr < 2\sqrt{\dr}. \] 

For $\mu<0$, all initial conditions in $R$ are attracted to $\br$ which is in
$L$. All initial conditions in $L$ converge on to $L^*$. Therefore the fixed
point is the unique attractor. For $\mu>0$, all initial conditions in $L$ move
linearly towards $\bl$ which is in $R$, and all points in $R$ spiral towards
$R^*$. Therefore $R^*$ is the unique attractor.

{\em Spiral attractor to spiral attractor having the same sense of
rotation:} This occurs \[\mbox{if} \;\;\;\; 0<\tl < 2\sqrt{\dl}
\hspace{0.1in} \mbox{and} \hspace{0.1in} 0 < \tr < 2\sqrt{\dr} \]
\[\mbox{or} \;\;\;\; -2\sqrt{\dl} < \tl < 0 \hspace{0.1in} \mbox{and}
\hspace{0.1in} -2\sqrt{\dr} < \tr < 0. \]

If the spiralling orbits in $L$ and $R$ have the same sense, there is an
overall spiralling orbit converging on the fixed point. Therefore there is an
unique period-1 attractor for both $\mu<0$ and $\mu>0$.

{\em Regular attractor to regular attractor:} \[ 2\sqrt{\dl} < \tl
< (1+\dl)  \hspace{0.1in} \mbox{and} \hspace{0.1in} 2\sqrt{\dr} < \tr <
(1+\dr), \]

{\em Flip attractor to flip attractor:} \[ -2\sqrt{\dl} > \tl > -(1+\dl)
\hspace{0.1in} \mbox{and}
\hspace{0.1in} -2\sqrt{\dr} > \tr > -(1+\dr), \]

{\em Regular attractor to flip attractor:} \[ 2\sqrt{\dl} < \tl <
(1+\dl)  \hspace{0.1in} \mbox{and} \hspace{0.1in} -2\sqrt{\dr} > \tr >
-(1+\dr). \]

In the above three cases, for $\mu<0$, initial conditions in $R$ move linearly
to $\br$. Since there must be a heteroclinic intersection of the stable
manifolds, all initial conditions converge on $L^*$. The situation for $\mu>0$
is similar.

\noindent{\bf Case 4:} In the following cases there can be bifurcation
from
multiple attractors to multiple attractors. There are general mechanisms
for the occurrence of coexisting attractors. 

{\em Spiral attractor to spiral attractor with opposite sense of
rotation:} This occurs \[ \mbox{if} \hspace{0.1in} 0 < \tl <2\sqrt{\dl}
\hspace{0.1in} \mbox{and} \hspace{0.1in} -2\sqrt{\dr} < \tr < 0 \] \[
\mbox{or} \hspace{0.1in} -2\sqrt{\dl}<\tl <0 \hspace{0.1in} \mbox{and}
\hspace{0.1in} 0<\tr < 2\sqrt{\dr}. \]

{\em Spiral attractor to flip attractor:} This occurs \[ \mbox{if}
\hspace{0.1in} -2\sqrt{\dl} < \tl < 2\sqrt{\dl} \hspace{0.1in} \mbox{and}
\hspace{0.1in} -2\sqrt{\dr} > \tr > -(1+\dr). \]

There can be multiple attractors on both sides of $\mu$, one of which is a
fixed point.

\noindent {\bf Case 5:} In the parameter space region \[ \tr < -(1+\dr) 
\hspace{0.1in} \mbox{and} \hspace{0.1in} \tl < 0, \] initial conditions in
$L$ move to $R$ and vice versa. Therefore the dynamics is governed by the
stability of the second iterate with one point in $L$ and the other in
$R$. 

The eigenvalues of the second iterate are 
\[
 \frac{1}{2}\left(
\tau_{{L}}\tau_{{R}}- \delta_{{R}}- \delta_{{L}}\right. \pm \]
\[\left.
\sqrt
{{\tau_{{L}}}^{2}{\tau_{{R}}}^{2}-2\,\tau_{{L}}\tau_{{R}}\delta_
{{R}}-2\,\tau_{{L}}\delta_{{L}}\tau_{{R}}+{\delta_{{R}}}^{2}-2\,\delta   
_{{R}}\delta_{{L}}+{\delta_{{L}}}^{2}} \right)\]

From this, the condition of stability of the period-2 orbit is obtained as
\begin{eqnarray}
 1-\tl\tr+\dl+\dr+\dl\dr>0 &\;\;\; \mbox{for}& \lambda_1<+1
\label{period2a} \\
 1+\tl\tr-\dl-\dr+\dl\dr>0 &\;\;\; \mbox{for} & \lambda_2>-1
\label{period2b}
\end{eqnarray}

\begin{figure*} [t]
\begin{center}
\epsfig{figure=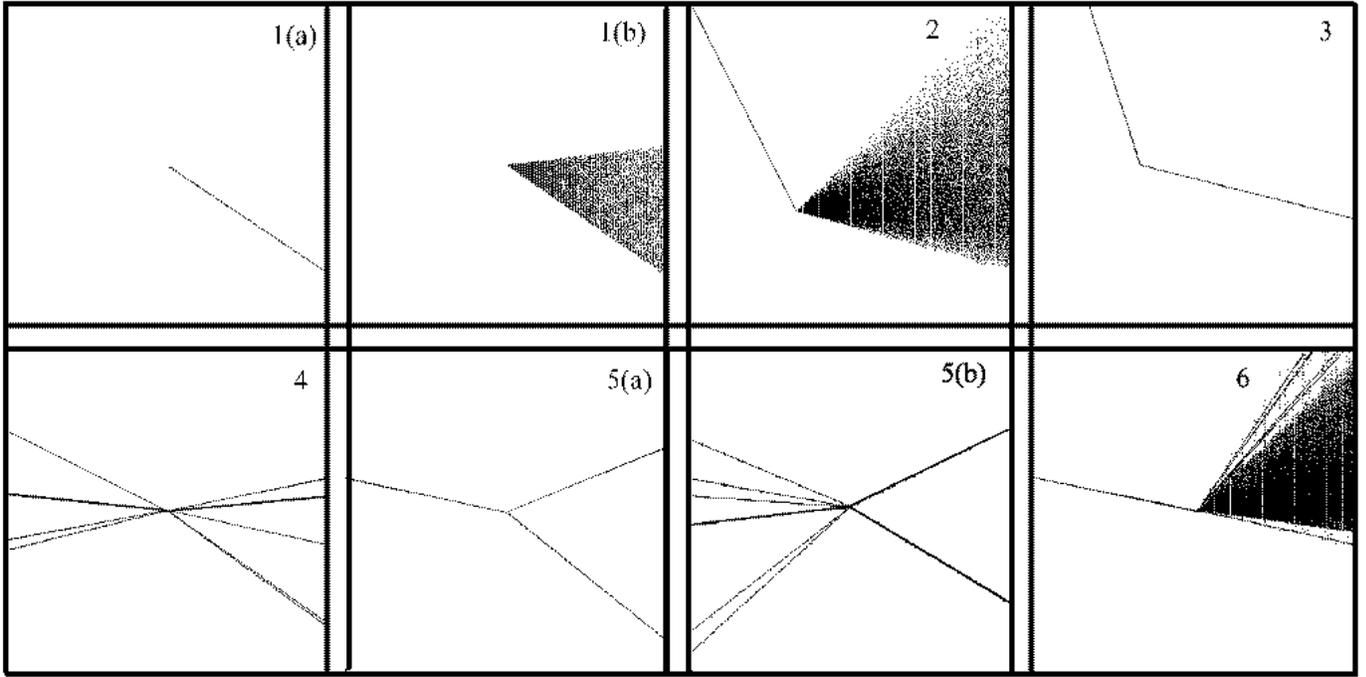,width=3.6in,angle=-90}
\end{center}
\caption{Representative
bifurcation diagrams of the normal form when $\mu$ is varied from a
negative value to a positive value. For the cases where multiple
attractors can exist, only one of many possibilities is shown. {\em
Case~1(a):} No attractor to period-1 attractor; {\em Case 1(b):} No
attractor to chaos; {\em Case 2:} Period-1 to chaos; {\em Case 3:}
Period-1 to period-1; {\em Case 4:} Period-1 + period-3 coexisting
attractors to period-1 + period-4 coexisting attractors; {\em Case 5(a):}
Period-1 to period-2; {\em Case 5(b):} Period-1 + period-11 coexsting 
attractors to period-2 attractor; {\em Case 6:} Period-1 to coexisting
period-5 + chaotic attractors.  } \label{bifs} \end{figure*}

There are three sub-cases:

\noindent {\bf Case~5(a):} If
\begin{eqnarray*}
& \tr < -(1+\dr) \hspace{0.1in}
\mbox{and} \hspace{0.1in}    \tl < -2\sqrt{\dl} \\
& \mbox{and} \hspace{0.1in} 1-\tl\tr+\dl+\dr+\dl\dr>0, \end{eqnarray*} then
there is a unique period-1 attractor for $\mu<0$ and a unique period-2
attractor for $\mu>0$.

For $\mu<0$, $L^*$ is a flip attractor and $\bar{R}^*$ is a flip saddle. All
initial conditions in $L$ converge on $L^*$ and all initial conditions in $R$
go to $L$ in the first iteration and then converge on to $L^*$. For $\mu>0$,
the condition (\ref{period2a}) ensures the stability of the period-2
orbit. The
existence of heteroclinic intersection makes the attractor unique.

This is like a period doubling bifurcation occurring on the borderline. In
contrast with standard period doubling bifurcation, the distinctive feature of
the border collision period doubling is that as $\mu$ is varied through zero,
the bifurcated orbit does not emerge orthogonally from the orbit before the
bifurcation.

\noindent {\bf Case~5(b):} If
\begin{eqnarray*} & \tr <  -(1+\dr)  \hspace{0.1in}
\mbox{and} \hspace{0.1in} -2\sqrt{\dl} <  \tl < 0 \\
& \mbox{and} \hspace{0.1in} 1-\tl\tr+\dl+\dr+\dl\dr>0, \end{eqnarray*} then
for $\mu<0$ there can be multiple attractors one of which is a period-1 fixed
point. For $\mu>0$, the period-2 orbit involving both $L$ and $R$ is stable.
Therefore there is an unique period-2 attractor.

\noindent {\bf Case~5(c):} If
\begin{eqnarray*} & \tr <  -(1+\dr)  \hspace{0.1in}
\mbox{and} \hspace{0.1in} -(1+\dl) <   \tl  \\
& \mbox{and} \hspace{0.1in} 1-\tl\tr+\dl+\dr+\dl\dr<0, \end{eqnarray*} then
there is a period-1 attractor for $\mu<0$.
For $-(1+\dl)<\tl<-2\sqrt{\dl}$, the eigenvalues of $L^*$ are real and
coexisting attractors cannot occur. Otherwise multiple attractors may
exist.
  For $\mu>0$, since (\ref{period2a})
is not satisfied, it implies that the fixed point of the twice iterated map is
unstable. The eigenvalues are real and initial conditions diverge away
from it along the unstable eigenvector.
Therefore there can be no attractor for $\mu>0$.

\noindent {\bf Case~5(d):} If
\[  \tl<-(1+\dl)\hspace{0.1in} \mbox{and} \hspace{0.1in} \tr<-(1+\dr), \] 
there is no attractor for both positive and negative values of $\mu$ since all
the fixed points of the first and second iterate are unstable.

\noindent {\bf Case 6:}{\em Spiral attractor to flip saddle:} This occurs
\[
\mbox{if} \hspace{0.1in} 0 < \tl < 2\sqrt{\dl} \hspace{0.1in} \mbox{and}
\hspace{0.1in} \tr < -(1+\dr). \] For $\mu<0$, there can be multiple
attractors one of which is a period-1 fixed point. The asymptotic behavior
for $\mu\!>\!0$ may be a periodic attractor (of periodicity greater than
unity), or chaotic attractor.
As $\tl$ is increased, periodic windows of successively higher 
periodicities (2,3,4,...) occur, and there are windows of chaos between
two such periodic windows. The period-$n$ attractor comes into existence
through a border collision pair bifurcation in the $n$th iterate and goes
out of existence when the period-$n$ fixed point becomes unstable. From 
(\ref{period2b}), the
stability boundary of period-2 attractor is given by
$1+\tl\tr-\dl-\dr+\dl\dr=0$. For higher iterates such analytical
expressions for the boundary of periodic windows become involved and are
not presented here. There
is no mechanism to prevent the occurrence of multiple attractors.

This gives a complete description of the bifurcations that can occur at
various regions of the parameter space of the normal form (\ref{normal}).
Representative bifurcation diagrams of the cases (where attractors exist) are
shown in Fig.\ref{bifs}.

\begin{figure}[tbh]
\begin{center}
\epsfig{figure=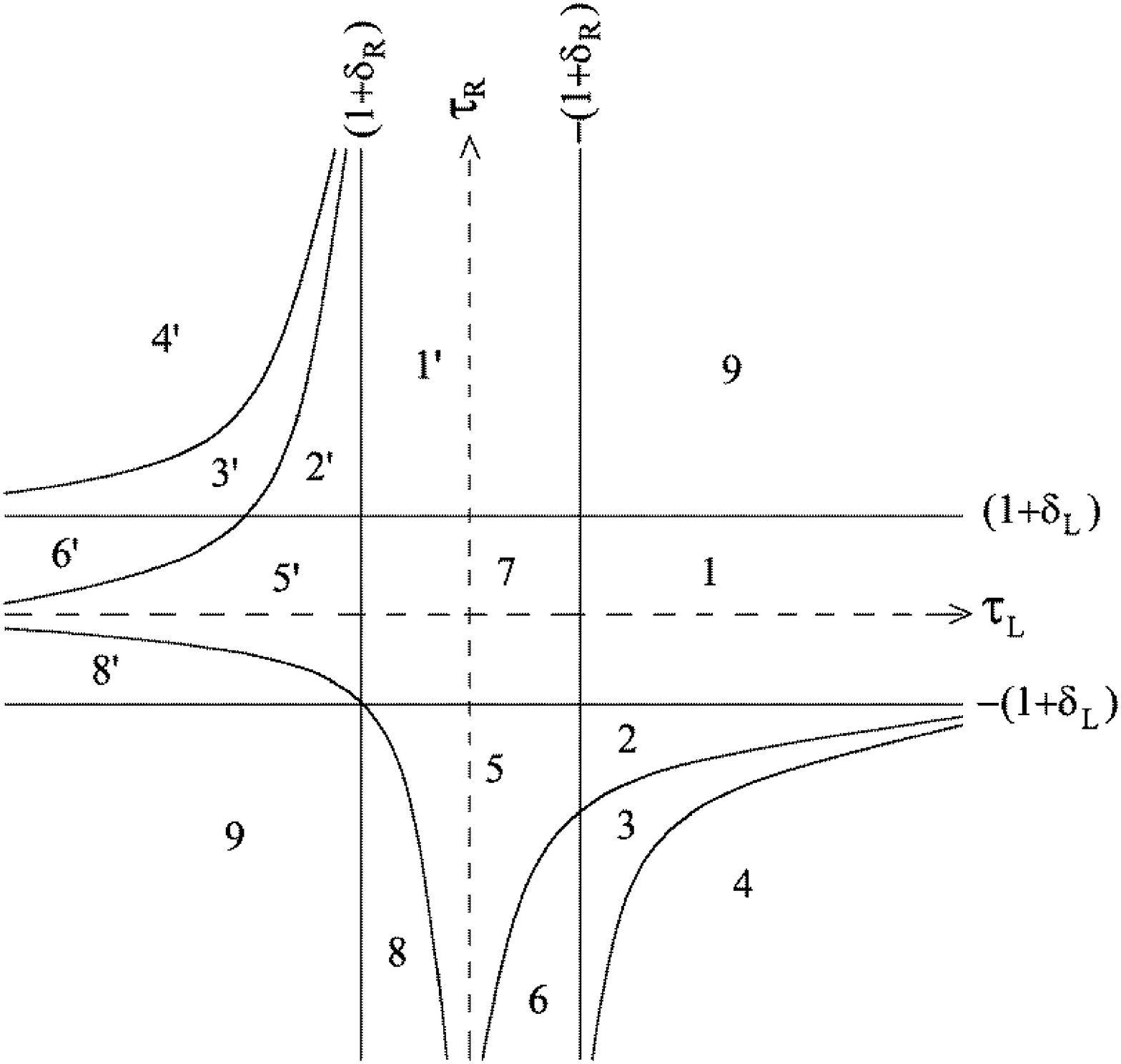,width=3.4in}
\end{center}
\caption{Schematic diagram of the parameter space partitioning for
$-1\!<\!\dl\!<\!0$ and
$-1\!<\!\dr\!<\!0$ into regions with the same qualitative
bifurcation phenomena. (1) No fixed point to period-1; (2) No fixed point
to period-2; (3) No fixed point to chaos; (4) No fixed point to unstable
chaotic orbit, no attractor; (5) Period-1 to period-2; (6) Period-1 to
chaos; (7) Period-1 to period-1; (8)
Period-1 to no attractor; (9) No
attractor to no attractor.
The regions shown in primed
numbers have the
same bifurcation behavior as the unprimed ones when $\mu$ is varied in the
opposite direction.}
\end{figure}

\subsection{The case of negative determinant}

If the determinant is negative, one has to find out which type of fixed point
changes to which type at it moves across the border. Depending on the type of
the fixed point at the two sides of the border, the bifurcations will be of
the same kind as discussed in the previous section.  For example, if $\dl ,
\dr < 0$ then the eigenvalues are real for all
values of $\tl$ and $\tr$.  Therefore there can be no coexisting attractors
anywhere in the parameter space. The region of stability of period-2
attractor, given by conditions (12) 
and (13), is much larger. Moreover there is a region of parameter space  
where a border collision pair bifurcation results in the creation of a
period-2 attractor since condition (13) is satisfied. The partitioning of
the parameter space for negative determinants is given in Fig.9.

There is however a difference in the equation for the boundary
crisis in border collision pair bifurcation. For $-1\!<\!\dr\!<\!0$, we have
$1\!>\!\lambda_{1R}\!>\!0$, $\lambda_{2R}\!<\!-1$, and $R^*$ is located above
the x-axis. A positive value of $\lambda_{1R}$ implies that $\uml$ converges
on $\umr$ from one side. If
\beq \frac{\lambda_{1L}-1}{\tl-1-\dl}>\frac{\lambda_{2R}-1}{\tr-1-\dr}
\label{whichpoint} \eeq 
then the intersection of $\uml$ with the x-axis remains the rightmost point of
the attractor and (\ref{cond}) still gives the parameter range for boundary
crisis. But if (\ref{whichpoint}) is not satisfied, the intersection of $\umr$
with the x-axis becomes the rightmost point of the attractor, and the
condition of existence of the chaotic attractor changes to
\beq
{\frac { \lambda_{2R}-1}{\tau_{{R}}-1-\delta_{{R}}}}<{\frac {
\delta_{{L}}\left (\tau_{{L}}-\delta_{{L}}- \lambda_{2L}\right )
}{\left (\tau_{{L}}-1-\delta_{{L}}\right )\left (\delta_{{R}}
\lambda_{2L}-\delta_{{L}}\tau_{{R}}\right )}}
\eeq

For $\dl\!<\!0$ and $\dr\!<\!0$, $L^*$ is below the x-axis and the same logic
as above applies. But if $\dl\!<\!0$ and $\dr\!>\!0$, the stable manifold of
$R^*$ has a negative eigenvalue and hence $\uml$ does not approach $\umr$ from
one side. Therefore, if (\ref{whichpoint}) is not satisfied, there is no
analytic condition for boundary crisis --- it has to be determined
numerically.

\section{Conclusions}

In this paper we have investigated the various types of border collision
bifurcations that can occur in piecewise smooth maps by deriving a piecewise
affine approximation of the map in the neighborhood of the border. We have
shown that there can be basically eleven different types of border collision
bifurcations, classified under six ``cases''. We have presented a partitioning
of the parameter space into regions where qualitatively different bifurcations
occur.

This body of knowledge helps us in explaining the bifurcations observed in
experimental and numerical investigations of switching circuits, some of
which have been presented in Sec.2. For example, the
experimental bifurcations of the
type seen in Fig.\ref{bif2} can occur in Case~2 and a part of Case~6. A
period doubling bifurcation of the type shown in Fig.\ref{bbif1} can occur
in the second iterate of the map if the parameters fall under Cases 5(a),
5(b)  and a part of Case~6 (coexisting attractors can not
be
observed in experimental bifurcation diagrams). The sudden appearance of a
chaotic attractor as in Fig.\ref{bbif3} can occur in border collision pair
bifurcation and can be categorized under Case 1(b). Note that this
bifurcation occurs in the third iterate while the period-1 attractor is
present, and therefore the resulting chaotic attractor is not robust. 

The theoretical problem dealt in this paper was posed by the recent
investigations in switching electrical circuits. But we believe that such
atypical bifurcations will be observed in other nonsmooth physical systems
also and the theory developed in this paper will help in understanding the
nonlinear phenomena and bifurcations in such systems.

\subsection*{Acknowledgements}

We acknowledge the discussions with Professor J. A. Yorke and Dr. G. H. 
Yuan in the initial stages of this work, and the help provided by Dr. D.
Kastha and Mr. S. Das in obtaining the experimental results. This work was
partly supported by NSF-CNPq, USA and by III~4(23)/94-ET, DST, India.

\end{document}